\documentclass[preprint, resetfootnote]{aastex701}

\usepackage{soul}
\usepackage{CJK}
\usepackage[version=4]{mhchem}
\usepackage{siunitx}
\DeclareSIUnit\au{au}
\newcommand{\tnm}[1]{\,\tablenotemark{#1}}
\newcommand{\tnmspace}[1]{\,\tablenotemark{#1}}    
\newcommand{\tntext}[2]{\tablenotetext{#1}{\ \ #2}}

\begin{document}
\begin{CJK*}{UTF8}{gbsn}

\title{Interstellar Object 3I/ATLAS Observed from Mars by China's Tianwen-1 Spacecraft}

\shorttitle{3I/ATLAS Observed by Tianwen-1 Orbiter}

\correspondingauthor{Jian-Yang Li, Yan Geng, Jianjun Liu}

\author[orcid=0000-0002-2418-4495, sname=Ren, gname=Xin]{Xin Ren (任鑫)}
\affiliation{Key Laboratory of Lunar and Deep Space Exploration, National Astronomical Observatories, Chinese Academy of Sciences, Beijing, People's Republic of China}
\affiliation{University of Chinese Academy of Sciences, Beijing, People's Republic of China}
\email{renx@nao.cas.cn}

\author[orcid=0000-0002-7560-6707, gname=Wei, sname=Yan]{Wei Yan (严韦)}
\affiliation{Key Laboratory of Lunar and Deep Space Exploration, National Astronomical Observatories, Chinese Academy of Sciences, Beijing, People's Republic of China}
\email{yanw@nao.cas.cn}

\author[orcid=0000-0003-4936-4959, sname=Zhao, gname=Ruining]{Ruining Zhao (赵瑞宁)}
\affiliation{CAS Key Laboratory of Optical Astronomy, National Astronomical Observatories, Chinese Academy of Sciences, Beijing 100101, China}
\email{rnzhao@nao.cas.cn}

\author[orcid=0000-0003-4489-9794,sname=Wang, gname=Shu]{Shu Wang (王舒)}
\affiliation{CAS Key Laboratory of Optical Astronomy, National Astronomical Observatories, Chinese Academy of Sciences, Beijing 100101, China}
\email{shuwang@nao.cas.cn}

\author[sname=Gao, gname=Xingye]{Xingye Gao (高兴烨)}
\affiliation{Key Laboratory of Lunar and Deep Space Exploration, National Astronomical Observatories, Chinese Academy of Sciences, Beijing, People's Republic of China}
\email{gaoxy@nao.cas.cn}

\author[gname=Qiang, sname=Fu]{Qiang Fu (付强)}
\affiliation{Key Laboratory of Lunar and Deep Space Exploration, National Astronomical Observatories, Chinese Academy of Sciences, Beijing, People's Republic of China}
\email{fuq@nao.cas.cn}

\author[orcid=0000-0003-2263-5165, sname=Zhang, gname=Qing]{Qing Zhang (张庆)}
\affiliation{Planetary Environmental and Astrobiological Research Laboratory (PEARL), School of Atmospheric Sciences, Sun Yat-sen University, Zhuhai, Guangdong, People's Republic of China}
\email{zhangq735@mail.sysu.edu.cn}

\author[orcid=0000-0002-5033-9593, sname=Yang, gname=Bin]{Bin Yang (杨彬)}
\affiliation{Instituto de Estudios Astrof\'{i}sicos, Facultad de Ingenier\'{i}a y Ciencias, Universidad Diego Portales, Santiago, Chile}
\email{bin.yang@mail.udp.cl}

\author[orcid=0000-0001-9067-7477, sname=Hui, gname=Man-To]{Man-To Hui ({\CJKfamily{bsmi}許文韜})}
\affiliation{Shanghai Astronomical Observatory, Chinese Academy of Sciences, Shanghai, People's Republic of China}
\email{mthui@shao.ac.cn}

\author[orcid=0000-0002-5026-6937, gname=Zhiyong, sname=Xiao]{Zhiyong Xiao (肖智勇)}
\affiliation{Planetary Environmental and Astrobiological Research Laboratory (PEARL), School of Atmospheric Sciences, Sun Yat-sen University, Zhuhai, Guangdong, People's Republic of China}
\affiliation{State Key Laboratory of Lunar and Planetary Sciences, Macau University of Science and Technology, Taipa, Macau, People's Republic of China}
\email{xiaozhiyong@mail.sysu.edu.cn}

\author[orcid=0000-0001-9329-0315, sname=Liu, gname=Xiaodong]{Xiaodong Liu (刘晓东)}
\affiliation{School of Aeronautics and Astronautics, Shenzhen Campus of Sun Yat-sen University, Shenzhen, Guangdong, People's Republic of China}
\affiliation{Shenzhen Key Laboratory of Intelligent Microsatellite Constellation, Shenzhen Campus of Sun Yat-sen University, Shenzhen, Guangdong, People's Republic of China}
\email{liuxd36@mail.sysu.edu.cn}

\author[gname=Cunhui, sname=Li]{Cunhui Li (李存惠)}
\affiliation{National Key Laboratory of Materials Behavior and Evaluation Technology in Space Environments, Lanzhou Institute of Physics, Lanzhou, Gansu, People's Republic of China}
\email{licunhui@spacechina.com}

\author[gname=Renhao, sname=Tian]{Renhao Tian (田仁浩)}
\affiliation{Key Laboratory of Lunar and Deep Space Exploration, National Astronomical Observatories, Chinese Academy of Sciences, Beijing, People's Republic of China}
\email{tianrh@bao.ac.cn}

\author[gname=Wenguang, sname=Liu]{Wenguang Liu (刘文光)}
\affiliation{Changchun Institute of Optics, Fine Mechanics and Physics, Chinese Academy of Sciences, Changchun, People's Republic of China}
\email{liuwenguang@ciomp.ac.cn}

\author[gname=Dong, sname=Wang]{Dong Wang (王栋)}
\affiliation{Changchun Institute of Optics, Fine Mechanics and Physics, Chinese Academy of Sciences, Changchun, People's Republic of China}
\email{wangd@ciomp.ac.cn}

\author[orcid=0009-0004-9570-8980,gname=Shaoran, sname=Liu]{Shaoran Liu (刘少然)}
\affiliation{Beijing Aerospace Command and Control Center, Beijing, People's Republic of China}
\email{30710455@qq.com}

\author[orcid=0009-0002-6651-519X,gname=Cong, sname=Ren]{Cong Ren (任聪)}
\affiliation{Beijing Aerospace Command and Control Center, Beijing, People's Republic of China}
\email{renc12590@163.com}

\author[gname=Jie, sname=Dong]{Jie Dong (董捷)}
\affiliation{Beijing Institute of Spacecraft System Engineering, Beijing, People's Republic of China}
\email{donghn13@163.com}

\author[gname=Xinbo, sname=Zhu]{Xinbo Zhu (朱新波)}
\affiliation{Shanghai Institute of Satellite Engineering，Shanghai, People's Republic of China}
\email{xinberg@163.com}

\author[gname=Pan, sname=Xie]{Pan Xie (谢攀)}
\affiliation{Shanghai Institute of Satellite Engineering，Shanghai, People's Republic of China}
\email{wakexie@163.com}

\author[orcid=0000-0003-3841-9977,gname=Jian-Yang, sname=Li]{Jian-Yang Li (李荐扬)}
\affiliation{Planetary Environmental and Astrobiological Research Laboratory (PEARL), School of Atmospheric Sciences, Sun Yat-sen University, Zhuhai, Guangdong, People's Republic of China}
\email[show]{lijianyang@mail.sysu.edu.cn}

\author[gname=Yan, sname=Geng]{Yan Geng (耿言)}
\affiliation{Lunar Exploration and Space Engineering Center, Beijing, People's Republic of China}
\email[show]{gengyan8015@163.com}

\author[orcid=0000-0002-9328-6532, gname=Jianjun, sname=Liu]{Jianjun Liu (刘建军)}
\affiliation{Key Laboratory of Lunar and Deep Space Exploration, National Astronomical Observatories, Chinese Academy of Sciences, Beijing, People's Republic of China}
\affiliation{University of Chinese Academy of Sciences, Beijing, People's Republic of China}
\email[show]{liujj@nao.cas.cn}

\shortauthors{Ren et al.}

\submitjournal{\apjl}
\accepted{\today}

\begin{abstract}

China's Tianwen-1 Mars orbiter successfully imaged the third interstellar object, 3I/ATLAS, during its close encounter with Mars using the onboard HiRIC CMOS camera. This is China's first deep-space observation of an astronomical object. These observations constitute the first imaging of this object from a vantage point significantly out of its orbital plane, providing a unique constraint on dust dynamics.  Three observing epochs between 2025 September 30 and October 3 reveal clear changes in coma and tail morphology driven by the rapidly evolving viewing geometry.  Comparison with Finson-Probstein dust dynamical models indicates that the coma is dominated by large grains with solar radiation pressure parameter $\beta \approx \num{e-3} \text{--} \num{e-2}$, corresponding to grain sizes of a few 100\,s \si{\um}.  The extent of the sunward coma implies dust ejection velocities of $3 \text{--} \SI{10}{m.s^{-1}}$. Despite the morphological evolution, the azimuthally averaged surface brightness profile remains nearly unchanged through the three epochs, transitioning from a radial slope near -1 close to the nucleus to slightly steeper than -1.5 at larger cometocentric distances, consistent with steady-state dust outflow accelerated by solar radiation pressure.  Photometry yields an average $Af\rho \sim \SI{2.0(0.2)e4}{cm}$ and a corresponding dust mass loss rate of $\dot{M}\sim\SI{e3}{kg.s^{-1}}$.

\end{abstract}


\keywords{\uat{Interstellar objects}{52} -- \uat{Comets}{280} -- \uat{Comet dust tail}{2312} -- \uat{Comae}{271} -- \uat{Comet origins}{2203}}


\section{Introduction}

Since its discovery on 2025 July 1 \citep{2025MPEC....N...12D}, the third interstellar object 3I/ATLAS (3I hereafter) has been extensively observed and studied using ground- and space-based observing facilities, including prediscovery archival data \citep[e.g.,][]{2025MNRAS.542L.139B, 2025RNAAS...9..266F, 2025ApJ...991L...2F, ye_prediscovery_2025}.  Its high orbital eccentricity of 6.14 indicates a dynamical age of multi-billion years, much longer than those of the previous two interstellar objects \citep{taylor_kinematic_2025}, highlighting its uniqueness among the only three interstellar objects discovered to date.

3I shows clear cometary activity with a coma with both dust and gas observed. Spectroscopic observations from the James Webb Space Telescope (JWST) in early August 2025 revealed a much higher \ce{CO2} to water ratio than almost all solar system comets, together with the detection of \ce{CO} and water vapor \citep{cordiner_jwst_2025}.  Water sublimation started to increase in August 2025 when the comet moved inside \SI{2.5}{\au} \citep{xing_water_2025, 2025ATel17515....1J, 2025CBET.5625....1C}.  Water ice was also detected based on the characteristic spectral absorption at \SI{2}{\um}, suggesting that the icy grains were ejected from the nucleus into the coma, likely by \ce{CO2} sublimation \citep{yang_spectroscopic_2025, cordiner_jwst_2025, 2025RNAAS...9..242L}.  The post-perihelion JWST observations also revealed other species such as \ce{CH4} \citep{2026arXiv260122034B}.

Some physical properties of 3I's dust coma and the nucleus have been determined from the images collected by the Hubble Space Telescope (HST) and ground-based facilities.  Early observations suggested a nuclear radius between \num{0.22} and \SI{2.8}{km} \citep{jewitt_hubble_2025}, which was later revised to \SI{1.3(0.2)}{km} \citep{2026arXiv260121569H}.  A rotational period of the nucleus of around 16 hours was claimed by \citet{santana-ros_temporal_2025}.  The slow ejection velocity of about \SI{5}{m.s^{-1}} for the coma dust indicates possibly large grains of tens to hundreds of microns in size, released at a mass loss rate $\dot{M}\sim \num{12} \text{--} \SI{120}{kg.s^{-1}}$ in late July 2025 \citep{jewitt_hubble_2025}.  The coma morphology observed in similar time showed a lack of dust tail in the anti-solar direction, but rather a sunward feature \citep{de_la_fuente_marcos_assessing_2025, jewitt_hubble_2025}.  Such an unusual morphology has been explained by the slow reaction of large grains to solar radiation pressure (SRP) since the onset of dust activity \citep{jewitt_hubble_2025}, which probably occurred in April 2025 when the comet was at $\sim$\SI{9}{\au} from the Sun \citep{ye_prediscovery_2025}. The coma morphology also revealed a possible jet towards the Sun with periodic wobbling in its projected direction, consistent with the reported nucleus spin period \citep{serra-ricart_pre-perihelion_2026}, although no detection was reported elsewhere.  Photometry suggests that 3I had a rapid increase of dust activity inbound towards perihelion \citep{ye_prediscovery_2025, 2026PASP..138a4403Z}.

\begin{figure*}[ht!]
\plotone{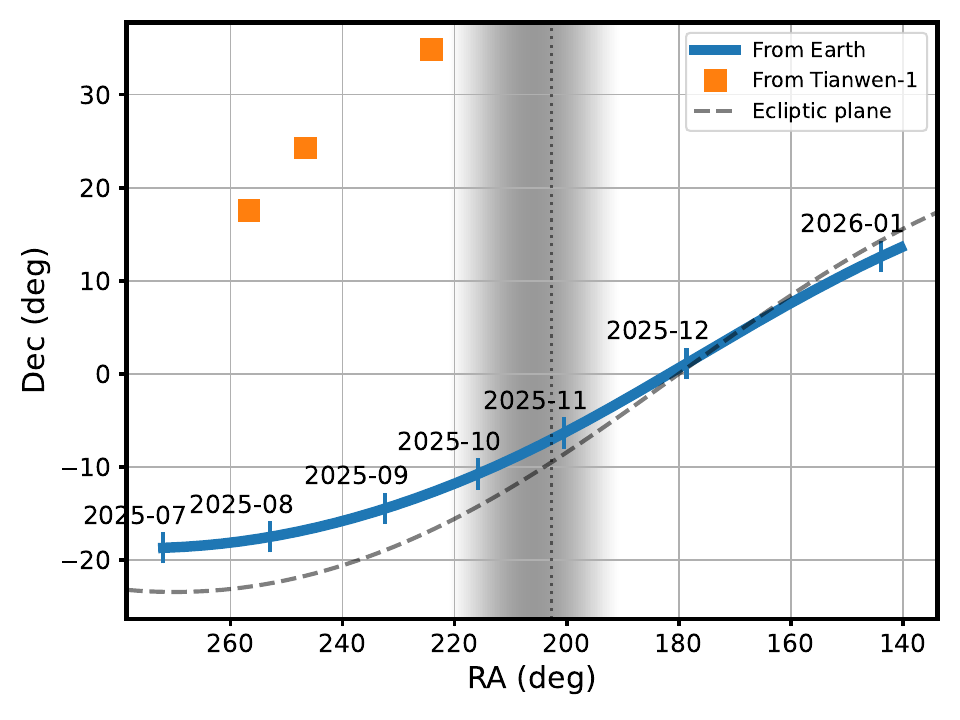}
\caption{The sky coordinate of 3I from Earth between 2025 July 1 and early 2026 January (blue line) and from Mars during the three epochs of Tianwen-1 observations (filled orange squares). The thin dashed curve marks the ecliptic plane. The vertical dotted line marks the RA of 3I from Earth at perihelion, and the shaded area marks the range of RA with solar elongation \textless\SI{45}{\degree}. While no ground-based optical observations are possible near perihelion and solar conjunction, monitoring can still be continued with space-based facilities.
\label{fig:geom}}
\end{figure*}

However, 3I had a solar conjunction around 2025 October 21 as it approached perihelion at \SI{1.36}{\au} on 2025 October 30, making it difficult or impossible to observe from the ground or near-Earth space.  Fortunately, it made a close approach to Mars at \SI{0.194}{\au} on 2025 October 3 UT 04:38, providing excellent observing conditions from Mars at solar elongations \textgreater\SI{125}{\degree} and a total magnitude brighter than \SI{8}{mag} to fill the observational gap. More importantly, the vantage point from Mars is the only opportunity to obtain observations from outside the orbital plane of 3I, providing a much better constraint on the dust grain size. This is because 3I has an orbital inclination of \SI{175}{\degree} and therefore, as shown in Fig.\,\ref{fig:geom}, it always locates close to the ecliptic plane.  Observations from ground or near-Earth space, or other spacecraft missions, all of which have low orbital inclinations, will be close to the orbital plane of 3I.  Whereas the \SI{0.2}{\au} close encounter with Mars provides an out-of-plane angle of \SI{35}{\degree} -- \SI{45}{\degree}.

In this work, we took advantage of the unique perspective from Mars and observed 3I during its Mars encounter using China's Tianwen-1 orbiter spacecraft, which was in a long elliptical orbit around Mars.  The Tianwen-1 orbiter was part of the orbiter-lander-rover combination of China's first planetary exploration mission, the Tianwen-1 mission. It performed orbital investigation of Mars and supported the science operations of the lander and the Zhurong rover on the surface of Mars \citep{2020NatAs...4..721W, 2021SSRv..217...26L, 2021AdSpR..67..812Z}.  As the first-ever attempt at observing an astronomical object using a planetary exploration spacecraft in China, the success demonstrated the capability and flexibility of the Tianwen-1 orbiter spacecraft, and set an example for exploiting operating spacecraft in deep space to perform target-of-opportunity astronomical observations.  Our data provided unique observational constraints on the dust particle size of 3I, as well as the constraints on other properties such as the dust expansion speed, morphology, and dust production during its Mars close approach.

\section{Observations and data reduction}

\subsection{Observations}

The Tianwen-1 orbiter has been operational for nearly five years after entering Mars orbit in 2021 February. The preparation of the 3I observations started in early September 2025. The optical payloads onboard the Tianwen-1 orbiter were designed to image the relatively bright Martian surface, which was \num{e4}--\num{e5} times brighter than 3I, and to capture a slow-moving scene. Therefore, the observations of 3I posed substantial challenges due to its large distance from Mars ($\sim$30 million km), high relative velocity ($\sim$\SI{86}{km.s^{-1}}), and the extreme faintness \citep[apparent magnitude \num{6.7}, c.f.][]{2025arXiv250815768E} compared to the Mars surface, requiring accurate pointing and a rigorous imaging strategy.

Considering all factors, the Tianwen-1 science operation team chose to use the CMOS array detector of the High-Resolution Imaging Camera \citep[HiRIC; see][]{meng_high_2021} to observe 3I. The CMOS camera was an auxiliary camera mounted on the focal plane of the HiRIC optical system. It has a 512$\times$512 pixel square CMOS array and a pixel scale of \SI{9.4}{\micro\radian} (\SI{1.94}{\arcsec}), resulting in a field of view of approximately \SI{0.28}{\degree}$\times$\SI{0.28}{\degree}. The CMOS camera was panchromatic, sensitive to the full visible wavelength range (\SI{400}{\nm}--\SI{1000}{\nm}). The images were digitized with 12-bit sampling, with a saturation level of \SI{4095}{DN}. The angular diameter of 3I's coma was predicted to be $\sim$\SI{3}{\arcmin} during its Mars encounter \citep{2025arXiv250815768E}, extending $\sim$90 pixels in the HiRIC CMOS array. 3I is expected to move one CMOS pixel in \SI{3.8}{\sec} in the sky, allowing for the longest available exposure time of the CMOS camera of \SI{1.34}{\sec}. With the above settings, the Tianwen-1 orbiter observed 3I in three epochs from 2025 September 30 to October 3. The duration of each epoch was limited to \SI{30}{\sec} due to the available hardware settings and the memory size. Nineteen images were acquired continuously at a fixed camera pointing during each epoch. The observing circumstances and the corresponding geometry of the three epochs were summarized in Table\,\ref{tab:geometry}.

\begin{deluxetable*}{lccccccccc}
\tablecaption{Summary of Tianwen-1 HiRIC observations of 3I and the corresponding observing geometry, photometric zero point, and total magnitude.\label{tab:geometry}}
\tablehead{
\colhead{UTC} &
\colhead{$r_{\mathrm{h}}$\tnm{a}} &
\colhead{$\Delta$\tnm{b}} &
\colhead{$\alpha$\tnm{c}} &
\colhead{Sun PA\tnm{d}} &
\colhead{Vel. PA\tnm{e}} &
\colhead{Pl. Ang.\tnm{f}} &
\colhead{TA\tnm{g}} &
\colhead{ZPT\tnm{h}} &
\colhead{$G$\tnm{j}} \\
\colhead{} &
\colhead{(\si{\au})} &
\colhead{(\si{\au})} &
\colhead{(\si{\deg})} &
\colhead{(\si{\deg})} &
\colhead{(\si{\deg})} &
\colhead{(\si{\deg})} &
\colhead{(\si{\deg})} &
\colhead{(\si{mag})} &
\colhead{(\si{mag})}
}
\startdata
2025-09-30T16:00 & 1.716 & 0.231 & 39.0 & 36.7  & 316.1 & -35.6 & 319 & \num{18.8(0.6)} & \num{7.5(0.6)} \\
2025-10-01T16:00 & 1.694 & 0.208 & 40.6 & 20.8  & 311.5 & -40.4 & 320 & \num{18.2(0.1)} & \num{6.7(0.1)} \\
2025-10-03T05:47 & 1.660 & 0.194 & 48.9 & 351.0 & 299.5 & -44.3 & 321 & \num{18.7(0.7)} & \num{7.0(0.7)} \\
\enddata
\tntext{a}{Heliocentric distance}
\tntext{b}{Observer–target distance}
\tntext{c}{Phase angle}
\tntext{d}{Position angle of the Sun–target vector}
\tntext{e}{Position angle of the target velocity vector}
\tntext{f}{Plane angle, with negative values meaning viewing from the south}
\tntext{g}{True anomaly}
\tntext{h}{Photometric zero point.  See \S\ref{subsec:reduction}}
\tntext{j}{Total magnitude in \SI{5000}{km} aperture in $G$ band. Here $G$ does not correspond to the standard Gaia G band, but rather to a white-light response. See \S\ref{subsec:reduction} and \S\ref{subsec:photometry} for details.}

\end{deluxetable*}

\subsection{Data reduction and calibration}\label{subsec:reduction}

A total of 57 images were successfully acquired with 3I easily visible in all images slowly drifting against the sky background within \SI{3}{\arcmin} of the frame center. The peak signal of the comet is between 1200 and \SI{2000}{DN}, which is about 1/3 of the saturation level and well above the bias level of $\sim$\SI{210}{DN}, indicating a peak signal-to-noise ratio (SNR) of $\gtrsim$\num{30}. The coma appears to be about \SI{90}{\arcsec} across.

Basic data reduction includes dark current and fixed-pattern noise removal, bias subtraction, and flatfielding.  All images displayed a fixed pattern of strip noise along the line direction at a level of $\lesssim$\SI{20}{DN}. To remove this noise, we stacked all images and used the minimum values of each pixels for the stack to construct a template.  Given the different star fields in three epochs and the different locations of 3I in all frames, this template could be subtracted from each frame to remove the dark current and fixed-pattern noise. Bias was removed using the median value of each individual frame after sigma-clipping. We then used the pre-flight flat-field data obtained in the laboratory to perform flat-field.

Photometric calibration was performed using field stars. Tens of field stars were identified in each frame to derive plate solutions via Astrometry.net \citep{Lang:2010}. For flux calibration, we adopted the Gaia $G$ catalog \citep{GaiaCollaboration:2023, GaiaCollaboration:2016}, assuming that its broad bandpass provides the best approximation for the unknown white-light transmission of our camera. This choice should minimize systematic errors relative to other narrower standard filters. We cross-matched the stars within a \SI{5}{\arcsec} radius of their corresponding catalog positions, and determined zero-points using a sigma-clipping regression to exclude outliers. The resulting zero-points are listed in Table\,\ref{tab:geometry}.

During each epoch, the pointing and orientation of the camera remained fixed, and 3I moved in the field of view against the background stars. Therefore, we calculated the projected direction of 3I's velocity vector with respect to the Tianwen-1 spacecraft in the sky plane and compared it with the direction of the apparent motion of 3I in the field of view for each epoch to determine the projected position angle of the J2000 north in the images. The parallax due to spacecraft motion during each 30-second observation epoch is about \SI{0.24}{\arcsec}, and the parallax with respect to the Mars center is up to \SI{100}{\arcsec}. Therefore, the uncertainty in the image orientation introduced by any uncertainty in the spacecraft position during the observations is negligible for our analysis.

\begin{figure*}[ht!]
\plotone{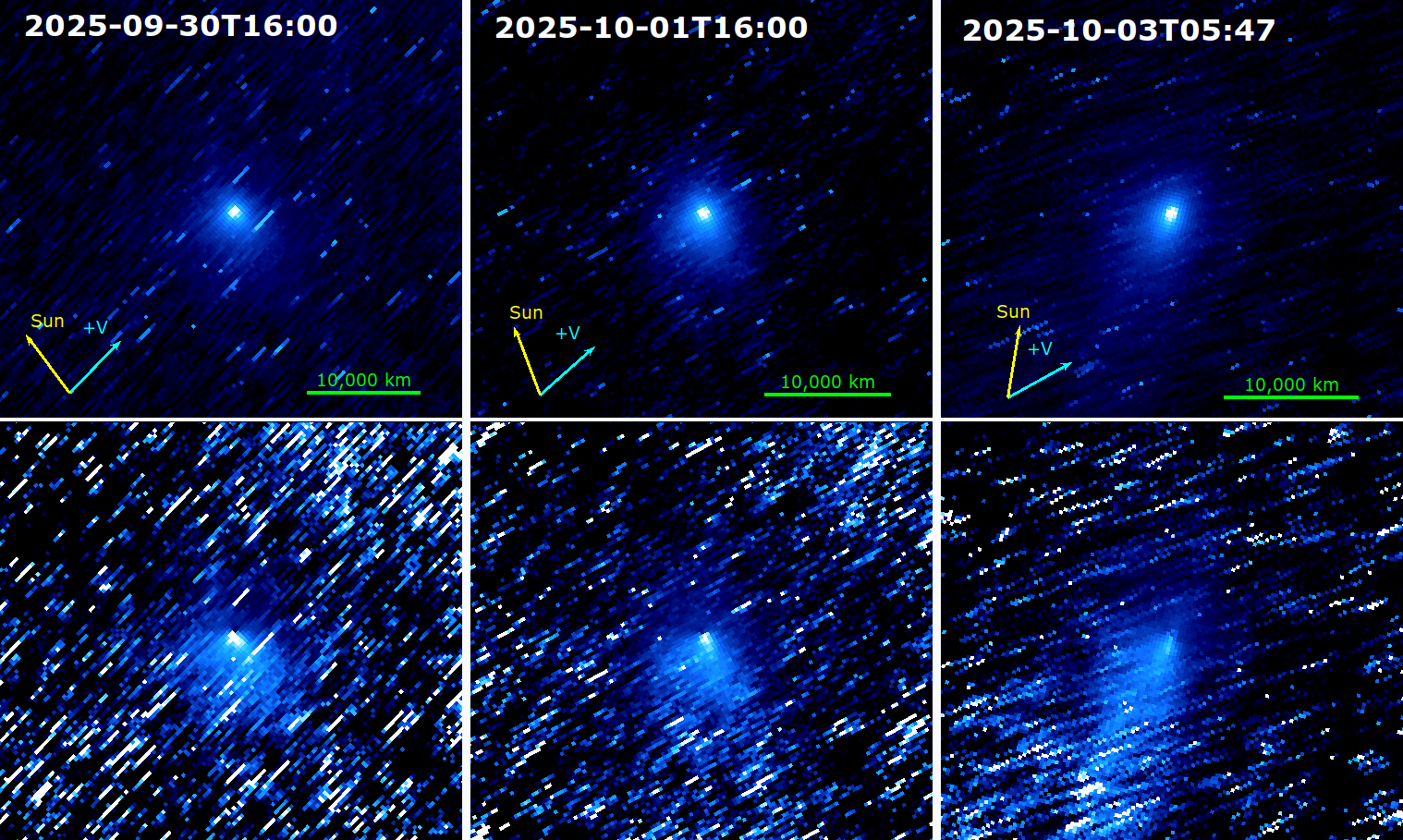}
\caption{Stacked images of 3I/ATLAS acquired by HiRIC CMOS camera onboard Tianwen-1 spacecraft (upper row) and the corresponding 1/$\rho$ divided images (lower row) from the three epochs.  All images are displayed north up and east to the left. The stacked images are displayed with logarithmic brightness stretch, and the enhanced images are displayed with a linear brightness stretch.  The arrows in the upper panels mark the projected directions of the Sun and the heliocentric velocity vector. The scalebars are \SI{10000}{km}.  The bright star trails are visible in each stack due to the apparent movement of 3I in the sky background during each 30-second observation epoch.
\label{fig:image}}
\end{figure*}

We then median-stacked all 19 images in each epoch centered at 3I to improve the SNR and remove bad pixels. The centroid of each stacked image was determined by a 2-D Gaussian fit to a 5$\times$5 pixel region centered at the brightest pixel of the comet. In order to facilitate morphological analysis and search for any jet features in the coma, we enhanced the stacked images by dividing a canonical 1/$\rho$ coma brightness model \citep{2014Icar..239..168S}. The SNR level of the images and the background stars does not warrant more sensitive enhancement techniques such as the Laplacian filter or the Larson-Sekanina filter. Our stacked images and enhanced images are shown in Fig.\,\ref{fig:image}.

\section{Results}

\subsection{Dust size and velocity}\label{subsec:size}

Given the wide bandpass of the HiRIC CMOS camera in the visible range, we can assume that the imaged coma of 3I is dominated by dust. Throughout the three epochs, the morphology of the dust coma and tail changed from a wide fan shape centered on position angle (PA) $\sim$\SI{180}{\degree} on September 30 and October 1 to a narrower, curved shape centered on PA$\sim$\SI{140}{\degree} (Fig.\,\ref{fig:image}). This is due to the change of viewing angle from Tianwen-1 by \SI{11.6}{\degree} and \SI{33.7}{\degree} from the first epoch to the second and third, respectively (Fig.\,\ref{fig:geom}). The plane angle also increased from \SI{36}{\degree} to \SI{44}{\degree} south of the orbital plane (Table\,\ref{tab:geometry}). The relatively high plane angle of the Tianwen-1 observations allows dust of different sizes to spread out in the sky plane under the influence of SRP. Therefore, these data enable us to determine the dust grain size in 3I's coma using a simple dust dynamical model.

\begin{figure*}[ht!]
\plotone{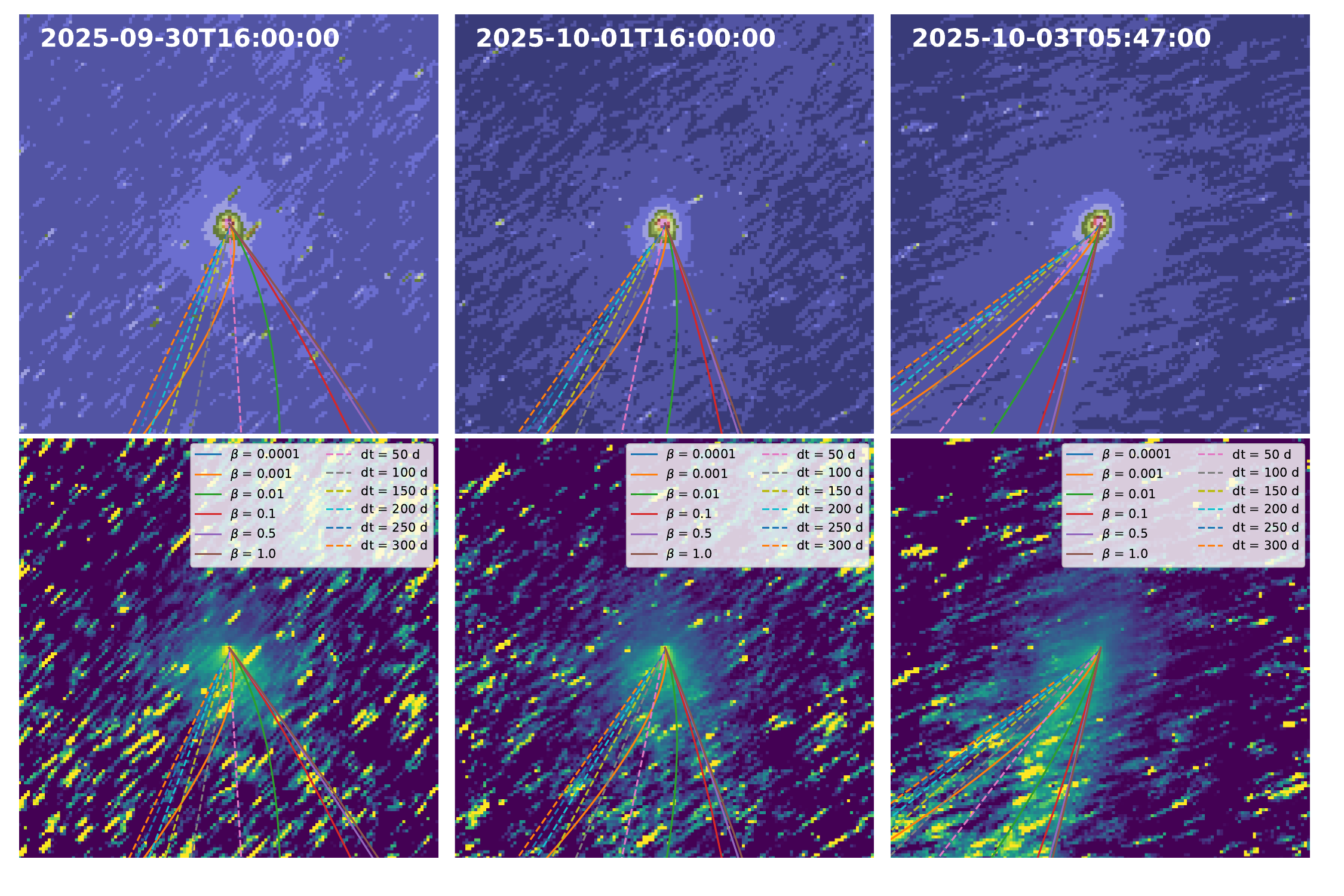}
\caption{The syndynes (solid curves) and synchrones (dashed lines) system of 3I/ATLAS corresponding to Tianwen-1 observations overlaid with the stacked images (upper row) and the 1/$\rho$ ratioed images (lower row). The image orientations and brightness stretch are the same as Fig.\,\ref{fig:image}, but displayed with different color tables. The syndynes correspond to $\beta = \num{1e-4}$ to 1, and the synchrones correspond to a release time of 50 -- 300 days before the observations, as shown in the legend in the lower row.
\label{fig:sync}}
\end{figure*}

The effect of SRP is characterized by the parameter $\beta$, which is defined as the ratio of SRP to solar gravity on a dust grain and is inversely proportional to grain size $r$. We calculated the zero-velocity syndynes and synchrones \citep{1968ApJ...154..327F} for a series of $\beta$ from \num{1e-4} to 1 and a series of dust release time from \num{300} to \SI{50}{days} before the observations for the three observing epochs (Fig.\,\ref{fig:sync}).

Compared to the syndyne system, the axes of symmetry of the coma and tail in the three epochs appear to align with $\beta = 0.01 \text{--} 0.001$, corresponding to grains of 100s of \si{\um} \citep{1979Icar...40....1B}. This is consistent with previous indications that the coma of 3I is composed of relatively large grains \citep{jewitt_hubble_2025}, but serves as a direct estimate from observations. Compared to the synchrone system, the images suggest that the observed coma is probably dominated by dust released within 2 -- 3 months before the observations, corresponding to after July when the comet entered about \SI{3}{\au} from the Sun. This is consistent with the dust being predominantly driven out by water sublimation. However, the synchrones of earlier than 100 days are close to each other, and it is impossible to accurately estimate the earliest dust release date.

The dust ejection velocity, $v_d$, can be inferred from the extent of the coma in the sunward direction, which represents the distance where the dust ejected sunward turns around under SRP. Following the simple model described by \citet{2005Icar..173..533F} and \citet{2013Icar..222..799M}, the quantity $v_d^2/\beta$ can be constrained by the coma standing-off distance, taking into account the various projection effects. From the sunward coma size of about \SI{10}{\arcsec} measured in the three epochs of the images (Fig.\,\ref{fig:profile}), we calculated $v_d=100~\beta^{1/2} ~\si{m.s^{-1}}$. Using $\beta = 0.01 \text{--} 0.001$ as previously estimated, $v_d$ is constrained to 3 -- \SI{10}{m.s^{-1}}, consistent with the previous estimate from HST images \citep{jewitt_hubble_2025}.

\subsection{Coma brightness distribution}

\begin{figure*}[ht!]
\plottwo{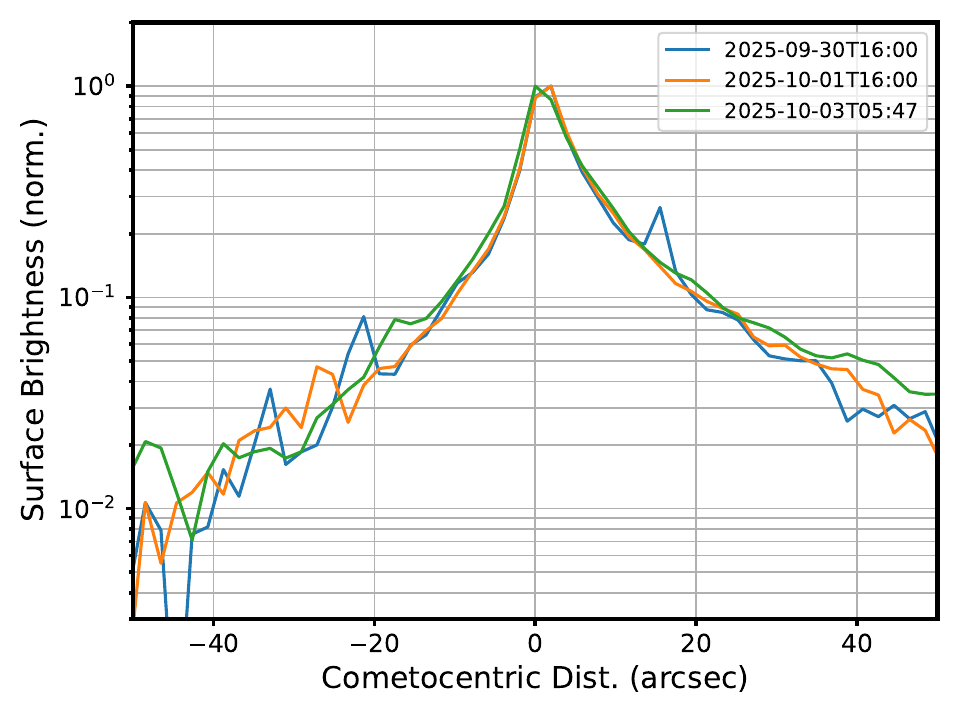}{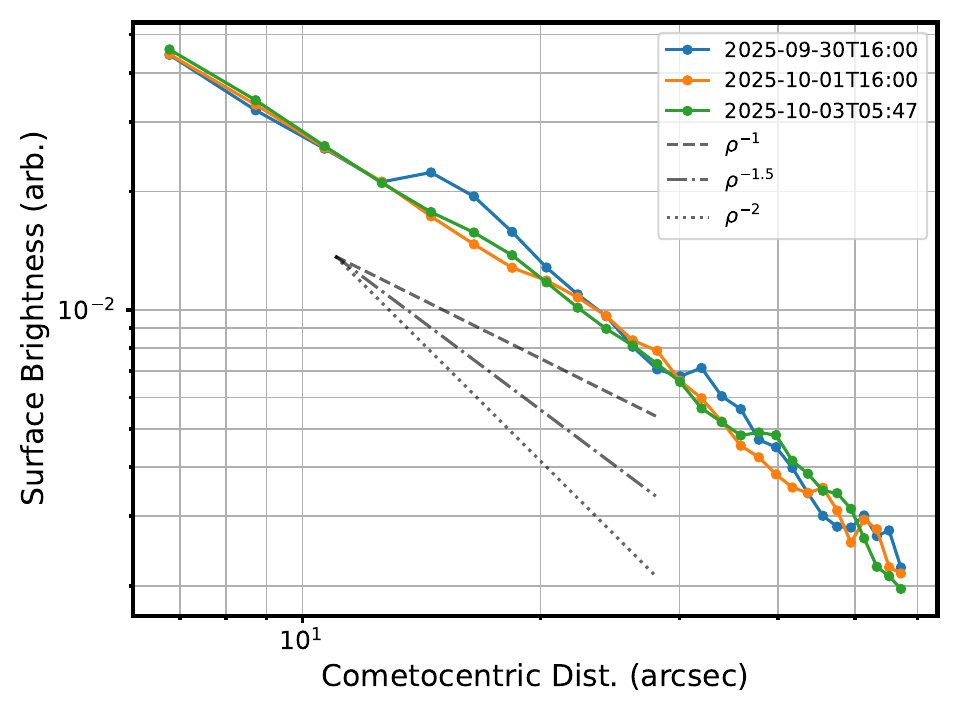}
\caption{(left) Surface brightness scan of the coma along the sun-antisun direction, normalized to the peak brightness. Left is the sunward direction, and right is the anti-sunward direction. The profile is averaged over a $\sim$\SI{10}{\arcsec} wide strip. An obvious asymmetry is shown between the sunward (left) and anti-sunward (right) direction, consistent with the effect of SRP. (right) Azimuthally averaged surface brightness profile with respect to cometocentric distance, derived from the photometry measured in a series of concentric annulus apertures centered on the centroid of the comet. Also shown are the lines representing exponential slopes of -1 and -2.
\label{fig:profile}}
\end{figure*}

Despite the different viewing angles causing changes in the overall morphology of the coma and tail across the three epochs, the brightness profile along the sun-antisun direction, as well as the azimuthally averaged surface brightness of 3I, remain almost the same (Fig.\,\ref{fig:profile}).  The surface brightness profile of 3I with respect to cometocentric distance exhibited a changing slope with distance: At $\lesssim\SI{20}{\arcsec}$ (\SI{1000}{km} -- \SI{3000}{km} ), the slope is close to -1; between $\sim$\SI{20}{\arcsec} and $\sim$\SI{30}{\arcsec} (\SI{4400}{km}), the slope increases to about -1.5; whereas at $\gtrsim\SI{30}{\arcsec}$ until the boundary of our measurement ($\sim$\SI{8000}{km}), the slope appears to be slightly steeper than -1.5 but with some uncertainty due to noise in the data at large cometocentric distances.  The average slope over the whole measured region within $\sim$\SI{55}{\arcsec} is -1.44. Note that the range of the physical distance covered in our brightness profile corresponds to \SIrange{0.5}{3.7}{\arcsec} in the HST images acquired on 2025 July 21 when 3I was at \SI{3.8}{\au} from the Sun \citep{jewitt_hubble_2025}, and within \SIrange{0.6}{4.4}{\arcsec} as observed from the ground in late September. Our overall observed coma brightness profile is consistent with that in ground-based observations at a similar time \citep{Jewitt:2025}, although we covered a different range of cometocentric distances in the coma.

For an isotropic and steady outflow dust coma, the surface brightness should follow a power law slope of -1, whereas for dust accelerated by SRP, the slope approaches -1.5. In the July HST images, the slope appears to become steeper than -1.5 at $\sim$\SI{3000}{km}, and was interpreted as due possibly to the water ice grains in the coma, sublimating as they receded from the nucleus \citep{jewitt_hubble_2025}, a scenario supported by the detection of water ice in the coma \citep{yang_spectroscopic_2025, cordiner_jwst_2025}. In our data, the slope beyond $\sim$\SI{3000}{km} is not obviously steeper than -1.5, consistent with dust being accelerated by SRP but not the sublimation of water ice. Indeed, given the heliocentric distance of 3I of \SI{1.7}{\au} during our observations, water ice is neither expected \citep{2006Icar..180..473B} nor reported \citep{2026RNAAS..10...26L}.

\subsection{Photometry\label{subsec:photometry}}

We performed aperture photometry on the flux-calibrated images with a fixed physical aperture radius of $\rho=\SI{5000}{km}$, corresponding to angular scales of \SI{30}{\arcsec}--\SI{36}{\arcsec}. The apparent magnitudes of 3I are listed in Table\,\ref{tab:geometry}. 

To facilitate a direct comparison with other space- and ground-based measurements, those magnitudes were then reduced to heliocentric magnitudes. We first converted the $G'$ magnitudes to Johnson $V$ by assuming a solar-like color (i.e., $G_{\rm BP}-G_{\rm RP} = 0.82$; \citealt{2018MNRAS.479L.102C}), which corresponds to a transformation of $V\sim G+0.2$ as defined in the Gaia EDR3 document\footnote{\url{https://gea.esac.esa.int/archive/documentation/GEDR3/}}. We then corrected the magnitude to \SI{1}{\au} observer distance and zero degree phase angle using
\begin{equation}\label{eqn:H}
    H = V - 5\log(\Delta/{\rm au}) + 2.5\log[\Phi_{\rm HM}(\alpha/{\rm deg})]~~,
\end{equation}
where $H$ is the heliocentric magnitude and $\Phi_{\rm HM}(\alpha)$ is the Halley-Marcus phase function \citep{2011AJ....141..177S}. The heliocentric V-band magnitudes of 3I were derived as $H=9.8$, $9.2$, and $9.6$ for the three epochs, respectively.

\begin{figure*}[ht!]
\plotone{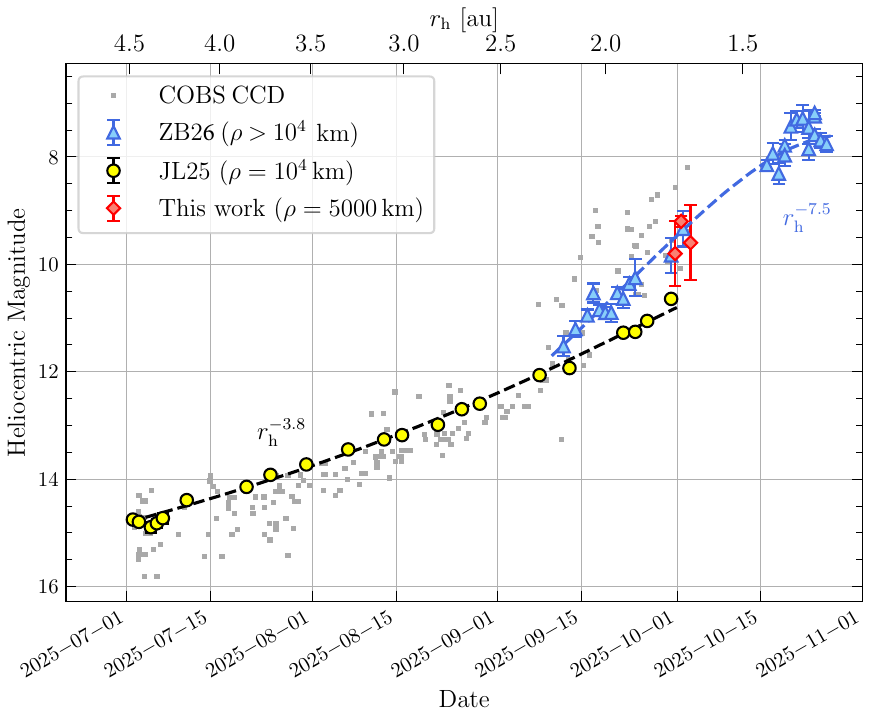}
\caption{Pre-perihelion heliocentric light curve of 3I/ATLAS. The measurements from this work (red squares, $\rho=\SI{5000}{km}$) are compared with space-based measurements from \citet[][blue triangles, $\rho>\SI{1e4}{km}$]{2026PASP..138a4403Z}, ground-based measurements from \citet[][yellow circles, $\rho=\SI{1e4}{km}$]{Jewitt:2025}, and CCD data from the COBS database (gray dots). Dashed lines indicate the brightening trends, which follow an $r_{\rm h}^{-3.8}$ power law at larger distances before transitioning to a steeper $r_{\rm h}^{-7.5}$ slope near $r_{\rm h} \sim \SI{2.5}{\au}$.\label{fig:lc}}
\end{figure*}

Fig.\,\ref{fig:lc} puts our results in the context of previously reported measurements from the ground \citep[][hereafter JL25]{Jewitt:2025}, the COBS database\footnote{\url{https://cobs.si/}, Credit: COBS Comet Observation Database - CC BY-NA-SA 4.0}, as well as the space-based solar coronagraph and heliospheric imager data \citep[][hereafter ZB26]{2026PASP..138a4403Z}. All data have been reduced to heliocentric magnitudes, except for the ZB26 measurements, which were corrected only for observer distance, since their narrow phase-angle coverage ($\lesssim\SI{10}{\degree}$) would introduce a correction of no more than $\sim$\SI{0.3}{mag} \citep{2026PASP..138a4403Z}. 

The COBS data were obtained using a variety of filters and aperture sizes, which naturally introduce substantial scatter; yet, they remain valuable as a long-term brightness baseline owing to the relatively long temporal coverage. In particular, they suggest a pronounced steepening of the brightening trend inside $\sim2.3$\,au. 
A similar behavior is seen in the ZB26 measurements, despite the different filters and aperture size used in those observations \citep{2026PASP..138a4403Z}. A fit to the ZB26 data yields a slope of $r_{\rm h}^{-7.5\pm1.0}$ \citep{2026PASP..138a4403Z}. Our measurements are broadly consistent with the COBS and ZB26 data, and, in particular, fall on the same steep brightening trend.


This steep slope may signal a transition in the dominant activity, likely associated with the onset of significant water-ice sublimation \citep{2026PASP..138a4403Z}. This picture is consistent with the steep increase in the water production rate over a similar heliocentric distance range, which followed $Q_{\rm H_2O}\sim r_{\rm h}^{-5.9\pm0.8}$ \citep{2026ApJ...998L..22T}. However, by scaling the gas-emission spectrum from \citet{Zhao:2026} to match the gas production rates near our observing epochs \citep[taken from ][]{Hutsemekers:2026}, we find that gas emission contributes no more than 10\% of the total flux in either $G$ or $V$ band. We therefore suggest that, although the brightening is associated with increasing outgassing, the band flux remained dominated by dust, likely because stronger outgassing lifted a larger amount of dust rather than because gas emission itself contributed substantially to the broadband flux.

The JL25 measurements are broadly consistent with the COBS data beyond $\sim$\SI{2.3}{\au}, although they appear systematically brighter, likely because they were obtained through the $R$ band \citep{Jewitt:2025}. However, they do not show an equally pronounced steepening at smaller heliocentric distances and are somewhat fainter than the other measurements. Given the high photometric quality of the JL25 data, this difference should be noted with caution. More homogeneous and better-sampled long-term monitoring will be needed to determine the possible cause of such a discrepancy.



\subsection{Dust production rate}

First introduced by \citet{AHearn:1984}, the quantity $Af\rho$ (product of albedo, filling factor, and projected linear radius of the aperture) is widely used as a proxy for the production of cometary dust. Following equation,
\begin{equation}
    Af\rho=\frac{(2r_{\rm h})^2}{\rho}10^{0.4\times(V_{\odot}-H)}~~,
\end{equation}
and using the apparent $V$ magnitude of the Sun $V_\odot=-26.76$ \citep{2012ApJ...752....5R}, we found $Af\rho=\num{1.2(0.7)e4}$, $\num{2.1(0.2)e4}$, and $\SI{1.4(0.8)e4}{cm}$ for the three epochs, respectively, and a weighted average of $\overline{Af\rho}=\SI{2.0(0.2)e4}{cm}$.

Under the assumption of steady-state isotropic expansion of the coma, the dust mass loss rate can be derived from $Af\rho$ by
\begin{equation}
    \dot{M}_{\rm d}=\frac{\rho_{\rm d} v_{\rm d}\overline{a} }{3p_V}Af\rho~~,
\end{equation}
where $\rho_{\rm d}$ is the dust mass density, $v_{\rm d}$ the expansion velocity, $\overline{a}$ the average size, and $p_V$ the geometric albedo of coma grains \citep{Jewitt:1991,Fink:2012}. Assuming that $\rho_{\rm d}=\SI{1}{g.cm^{-3}}$, $p_V=0.04$, and taking $v_{\rm d}\sim\SIrange{3}{10}{m.s^{-1}}$ and $\bar{a}\sim\SI{100}{\um}$ from \S\ref{subsec:size}, we derived an average dust mass loss rate of $\overline{\dot{M}_{\rm d}}\sim (\numrange{0.5}{1.7}) \times10^3\,\si{kg.s^{-1}}$, or of order \SI{e3}{kg.s^{-1}}. This is higher than the September 30 value reported by \citet{Jewitt:2025} by a factor of a few, mainly reflecting the differences in the photometry.


\section{Discussions}

\subsection{Comparison with 2I/Borisov}

The first interstellar object, 1I/'Oumuamua did not exhibit any dust activity \citep{2017Natur.552..378M}, although non-gravitational acceleration was detected, indicating possible outgassing \citep{2018Natur.559..223M}. The second interstellar object, 2I/Borisov (2I hereafter), shows obvious cometary activity with both dust \citep[e.g.][etc.]{2020ApJ...888L..23J} and multiple gas species \citep[e.g.][etc.]{2020ApJ...889L..10M, 2020MNRAS.495.2053D}, especially its high abundance of CO relative to water compared to solar system comets \citep{2020NatAs...4..867B}. Here we focus on the comparison of the dust activity and relevant properties between 2I and 3I, as well as water production that is the likely driver of dust release within a heliocentric distance of \SI{2}{\au} (Table\,\ref{tab:dust_comparison}).

\begin{deluxetable}{lcccc}
\tablecaption{Comparisons between 2I/Borisov and 3I/ATLAS.\label{tab:dust_comparison}}
\tablehead{
\colhead{Object} &
\colhead{$R$\tablenotemark{a}} &
\colhead{$\dot{M}$\tablenotemark{b}} &
\colhead{$r_d$\tablenotemark{c}} &
\colhead{$v_d$\tablenotemark{d}} \\
\colhead{} &
\colhead{(\si{km})} &
\colhead{(\si{kg.s^{-1}})} &
\colhead{(\si{\um})} &
\colhead{(\si{m.s^{-1}})} 
}
\startdata
2I/Borisov & 0.2--0.5\tnm{f} & $\sim$50\tnm{g} & $\sim$\num{e2}\tnmspace{f,h} & 4--7\tnm{h} \\
3I/ATLAS  & \num{1.3(0.2)}\tnm{j} & $\sim$180\tnm{k} & $\sim$\num{e2}\tnmspace{l} & $\sim$5\tnm{m},\ $\sim$10\tnm{l} 
\enddata
\tntext{a}{Nuclear radius}
\tntext{b}{Mass loss rate near $r_h=\SI{2}{\au}$}
\tntext{c}{Dust grain radius}
\tntext{d}{Dust ejection velocity}
\tntext{e}{Water production rate near $r_h=\SI{2}{\au}$}
\tntext{f}{\citet{2020ApJ...888L..23J}}
\tntext{g}{\citet{2020MNRAS.495.2053D}}
\tntext{h}{\citet{2020AJ....160...92H}}
\tntext{j}{\citet{2026arXiv260121569H}}
\tntext{k}{\citet{Jewitt:2025}}
\tntext{l}{This work}
\tntext{m}{\citet{jewitt_hubble_2025}}
\end{deluxetable}

The nuclear size of 3I is 3--5 times larger than that of 2I, resulting in an order of magnitude difference in their surface areas. This is probably a major factor that their dust production rates differ by a factor of $\sim$3. On the other hand, the dust properties of both interstellar objects are similar, with relatively large grains of 100s\,\si{\um} in their coma compared to typical solar system comets, ejected at similar speeds of a few to \SI{10}{m.s^{-1}}. Regarding water production rate, 3I is one order of magnitude higher than that of 2I.

For dust ejected by outgassing, the drag force exerted on a dust grain of radius $r_d$ is proportional to its cross-section, which scales as $r_d^2$. Because acceleration $a$ is inversely proportional to the dust grain mass, which scales as $r_d^3$, we have $a \propto r_d^{-1}$. The acceleration of dust grains due to outgassing will stop at a specific distance $L$, where the gas density drops to some point. Thus, we have the terminal velocity $v_d^2 \propto a L$, leading to $v_d \propto r_d^{-1/2}$. The scaling factor depends on the particular driving gas and $L$, which is determined by the production rate of the dominant volatile species and its basic properties, such as the molecular weight, the outflow speed, and the corresponding photodissociation timescale. The similar grain sizes and ejection velocities between 2I and 3I, therefore, probably suggest that the dust ejection from both objects is driven by the same gas species with similar production rates.

\subsection{Dust jet}

A dust jet is first reported in the coma of 3I in 2025 July after image enhancement that wobbles in a small range around PA$\sim$\SI{280}{\degree} with a periodicity consistent with its spin period reported elsewhere based on lightcurve \citep{serra-ricart_pre-perihelion_2026}. Recently, jets have been reported again using the Larson-Sekanina filtering technique to enhance the HST images acquired in 2025 November and December \citep{2026arXiv260110860S, 2026arXiv260218512S}. We searched for jet structures in our images but found none. Here we offer some thoughts about the dust jets in the coma of 3I.

First, we are highly doubtful about the recent report of jets from the HST images after image enhancement \citep{2026arXiv260110860S, 2026arXiv260218512S}.  Those authors claim three jets separated by $\sim$\SI{120}{\degree} from each other, all straight with no signs of any curvature \citep[see Fig.\,4 in][]{2026arXiv260218512S}. While no scale bar is provided, we assume the images are displayed in the original HST/WFC3/UVIS pixel scale, and estimate that all jets have projected lengths of \textgreater\SI{10}{pixels} when the comet was at \SI{1.83}{\au} from the Earth, resulting in a linear scale of \textgreater\SI{530}{\km} at the comet. Even if we take the fastest dust expansion speed of \SI{10}{m.s^{-1}} as derived by \citet{jewitt_hubble_2025} and in our work, it will still take at least $\sim$\SI{15}{\hour} for the dust to travel the length of the jets. Considering the projection effect, the travel time should be longer. If the spin period is $\sim$\SI{7}{\hour} based on \citet{2026arXiv260110860S}, that would be at least two rotations. Therefore, even if one of the jets is polar, the other two should show curvatures or arc-like shapes, and presumably discontinuous due to the movement of the source regions moving around the rotational axis, and in and out of daylight. Such curved or arc-like jets have been previously reported in many comets, e.g., Hale-Bopp (C/1995 O1) \citep{2003A&A...403..757H} and Machholz (C/2004 Q2) \citep{2007AJ....133.2001F}, etc. In any case, the directions of the whole jet cannot simply ''jump around'' with time.  In our quick processing of the same 2026 November and December HST images using the traditional $1/\rho$ and azimuthal median enhancement techniques for comets \citep{2014Icar..239..168S}, we do not see evidence of jets.

With the changing observing perspective of 3I from Earth from 2025 July to December and the unique perspective from Tianwen-1 (Fig.\,\ref{fig:geom}), we can analyze the possibility of 3I's dust jet based on the observations published so far. Given that the jet reported in the August ground-based observations wobbled in a small range around PA$\sim$\SI{280}{\degree} (i.e., almost aligned with the orbital plane), if it were a polar jet, the pole would lie near the sky plane at that time.  Thus, such a jet should also appear in the Tianwen-1 images because 3I was at similar RAs as the August ground-based observations but viewed from the south by $\sim$\SI{40}{\degree} (Fig.\,\ref{fig:geom}). The length scale of the jet was reported to be $\sim$\SI{15000}{km}, which should extend to $\sim$\SI{100}{\arcsec} in our images.

On the other hand, the 2025 November 30 HST observations had the RA of 3I changed by $\sim$\SI{80}{\degree} from July, and those in December differed further by $\sim$\SI{15}{\degree}. Thus, the lack of a jet viewed from a different perspective near the orbital plane in the HST data, combined with the lack of a jet viewed from a different perspective in the North-South direction in our data, is not in favor of the existence of a jet. If we consider that the true anomaly of 3I changed from $\sim$\SI{280}{\degree} in August to \SI{320}{\degree} in our observations, and the phase angle changed from $\sim$\SI{17}{\degree} to $\sim$\SI{40}{\degree}, then it is also possible that the jet has shut off because the Sun has moved away from the polar region. But that would require the polar region to have already been in a low-Sun condition in August, which did not favor the formation of the jet reported earlier.

\subsection{Large grains}

Both 2I and 3I have relatively large grains of 100s\,\si{\um} in their comae. Although the number is too small to conclude on a statistical basis, the large grains common in both are still worth some discussion.

Recent isotopic ratio measurements of 3I suggest that this object may have formed in a cold environment with temperatures \textless\SI{30}{\K}, probably indicating a formation in the outer protoplanetary disk or interstellar molecular cloud \citep{2026arXiv260306911C, 2026arXiv260307026S, 2026arXiv260307187O}. This conclusion is consistent with the high \ce{CO2} abundance in 3I \citep{cordiner_jwst_2025}. Although no isotopic measurements or limits have been reported for 2I, due mainly to its relatively low brightness, its high \ce{CO} abundance \citep{2020NatAs...4..867B} is at least consistent with an origin in a cold environment.


We suggest that the dominance of large grains in the comae of both 2I and 3I, if representing the properties of the nuclei rather than a selection effect in the dust ejection process, is consistent with an origin in the outer protoplanetary disk. Theoretical work has suggested that during planet formation around a young star, the presence of a gap in the planetary disk, perhaps due to the rapid formation of gas giants, could act as a barrier to the infall of large grains, but small grains can penetrate through more efficiently \citep{rice_dust_2006}. Therefore, large grains could be enriched outside the giant planet formation zone, whereas small grains would be enriched inside. Observations of a young star, PDS\,70, in thermal infrared confirm the enrichment of amorphous silicate grains in the inner disk as a result of the preferred filtration of fine grains towards the inner disk \citep{jang_dust_2024}. If this applies to the formation zones of 2I and 3I in the outer planetary disk, then it would result in the dominance of large grains incorporated into the comet.



\section{Conclusions}

In this work, we imaged the interstellar object 3I/ATLAS in three epochs near its close encounter with Mars using China's Tianwen-1 Mars orbiter.  This dataset provides the first observations from outside the orbital plane of 3I, directly constraining the dust grain size in the coma.  Our main results are summarized as follows:

\begin{enumerate}

    \item Over the three epochs of Tianwen-1 observations from 2025 September 30 to October 3, 3I showed a dust tail changing from a fan shape to a more elongated curving shape, centered near the anti-sunward direction.  This morphological change was due to the change in the observing perspective of about \SI{34}{\degree} over the four days.  The coma extends to about \SI{10}{\arcsec} in the sunward direction.

    \item Seen with a line-of-sight to the south of the orbital plane of 3I by \SI{35}{\degree} -- \SI{45}{\degree} in our Tianwen-1 images, the dust grains of various sizes in the coma are sufficiently separated in the sky plane.  We used the synchrones and syndynes model to determine that the dust coma of 3I is dominated by grains of 100s\,\si{\um} in size.  The size of the coma in the sunward direction suggests an ejection velocity of dust grains $v_d\sim 3 \text{--} \SI{10}{m.s^{-1}}$.

    \item Despite the changing morphology of the coma, the azimuthally averaged surface brightness of the coma remained almost unchanged.  The radial coma brightness distribution is characterized by a power law slope close to -1 within a few thousand \si{km} from the nucleus, increasing with distance to slightly steeper than -1.5 at $\gtrsim\SI{10000}{km}$.  This behavior is grossly consistent with steady-state dust outflow accelerated by SRP.

    \item The total brightness of 3I measured from the Tianwen-1 data is consistent with its overall brightening trend towards perihelion observed by ground- and space-based solar probes.  The $Af\rho$ is estimated to be \SI{2.0(0.2)e4}{cm}.  Under the assumption of steady state dust outflow model, based on the grain size and velocity that we measured, and assuming a dust density of \SI{1}{g.cm^{-3}}, the dust mass loss rate of 3I is $\dot{M}\sim\SI{e3}{kg.s^{-1}}$.
    
\end{enumerate}

The successful Tianwen-1 observation of 3I demonstrates that China's deep-space spacecraft is capable of making flexible target-of-opportunity observations of distant astronomical objects. It also underscores the scientific values of utilizing space assets in the solar system to complement the observations of important targets from different vantage points or when they are unobservable from the ground.

\begin{acknowledgments}

We are extremely grateful to the scientists and engineers of the Tianwen-1 Mission Science Operation Teams.  Without their enormous and dedicated efforts to the design, planning, and execution of the observations of 3I, this work would have been impossible. J.-Y.L. is supported by the National Natural Science Foundation of China (Grant No. 42530203). We acknowledge with thanks the comet observations from the COBS Comet Observation Database contributed by observers worldwide and used in this research. We thank the anonymous reviewer for critical reading and constructive suggestions that help us improve this manuscript.

\end{acknowledgments}

\begin{contribution}

X. Ren led data curation and contributed to the analysis and writing of the paper. R. Zhao contributed to data curation, analysis, and writing. B. Yang and M.-T. Hui contributed to the methodology and validation, and reviewed and edited the paper. W. Yan, X. Gao, Q. Zhang, Q. Fu, R. Tian contributed to data curation and methodology. S. Wang contributed to the project supervision and reviewed the paper. Z. Xiao, X. Liu, and C. Li contributed to conceptualization. S. Liu, C. Ren, W. Liu, D. Wang, J. Dong, X. Zhu, P. Xie contributed to the design and execution of observations. J.-Y. Li led the conceptualization, data analysis, interpretation, and writing of the paper. J. Liu and Y. Geng led project administration for resources, contributed to conceptualization, and reviewed the paper.



\end{contribution}

%
\facilities{Tianwen-1 (HiRIC), COBS}

\software{astropy \citep{2013A&A...558A..33A, 2018AJ....156..123A, 2022ApJ...935..167A},
sbpy \citep{2019JOSS....4.1426M}
          }





\bibliography{tw1_3i}{}
\bibliographystyle{aasjournalv7}



\end{CJK*}

\end{document}